\documentclass[twocolumn,showpacs,prb]{revtex4-1}%
\usepackage{amsfonts}
\usepackage{amsmath}
\usepackage{amssymb}
\usepackage{subfigure}
\usepackage{graphicx}
\usepackage{txfonts}%
\providecommand{\U}[1]{\protect\rule{.1in}{.1in}}

\begin{document}

\title{Robust wavefront dislocations of Friedel oscillations in gapped graphene}
\author{Shu-Hui Zhang$^{1}$$^\S$}
\email{shuhuizhang@mail.buct.edu.cn}

\author{Jin Yang$^{2}$$^\S$}

\author{Ding-Fu Shao$^{3}$}
\author{Zhenhua Wu$^{4}$}
\author{Wen Yang$^{2}$}
\email{wenyang@csrc.ac.cn}

\affiliation{$^{1}$College of Mathematics and Physics, Beijing University of Chemical Technology, Beijing,
100029, China}
\affiliation{$^{2}$Beijing Computational Science Research Center, Beijing 100193, China}
\thanks{$^\S$These two authors contributed equally to this work.}
\affiliation{$^{3}$Department of Physics and Astronomy and Nebraska Center for Materials and Nanoscience, University of Nebraska, Lincoln, Nebraska 68588-0299, USA}
\affiliation{$^{4}$Key Laboratory of Microelectronics Devices and Integrated Technology, Institute of Microelectronics of Chinese Academy of Sciences, 100029, People's Republic of China}

\begin{abstract}
Friedel oscillation is a well-known wave phenomenon, which represents the
oscillatory response of electron waves to imperfection. By utilizing the
pseudospin-momentum locking in gapless graphene, two recent experiments
demonstrate the measurement of the topological Berry phase by corresponding to
the unique number of wavefront dislocations in Friedel oscillations. Here, we study the Friedel oscillations in gapped graphene, in which the
pseudospin-momentum locking is broken. Unusually, the wavefront
dislocations do occur as that in gapless graphene, which expects the immediate verification in the current experimental condition. The number of
wavefront dislocations is ascribed to the invariant pseudospin winding number
in gaped and gapless graphene. This study deepens the understanding of
correspondence between topological quantity and wavefront dislocations in
Friedel oscillations, and implies the possibility to observe the wavefront dislocations of Friedel oscillations in intrinsic gapped two-dimensional materials, e.g., transition metal dichalcogenides.

\end{abstract}
\maketitle



Since the seminal discovery of graphene\cite{NovoselovScience2004},
two-dimensional materials have attracted wide interest because of their novel
physics and great potential applications\cite{s41565-020-0724-3}. Usually,
two-dimensional materials have high mobility, in which the antiparticles move
ballistically and exhibit unconventional quantum tunneling and
interference\cite{nphys384,nphys1198,CheianovScience2007,BeenakkerPRL2009,WuPRL2011,RevModPhys.81.109}. One can intentionally add one
or two impurities to form the impurity-design system, this kind of system is
charming because it is easily handled theoretically and experimentally, and
then it can be regarded as the model system for the exploration of ballistic
physics\cite{SettnesPRL2014}. Experimentally, scanning tunneling spectroscopy
(STM) is a proper tool for impurity-design system\cite{RevModPhys.86.959}.
More interesting, the dimension of two-dimensional materials are very unique.
On one hand, the bare surface properties of two-dimensional materials are also the
bulk properties in contrast to three-dimensional materials. On the other hand,
different from the one-dimensional materials, two-dimension materials with
two-dimensional parameter space is enough to evolve the global topological
quantity\cite{2006.08556}. Therefore, surface-sensitive STM measurement is
promising to explore the topological physics of impurity-design two-dimensional materials.

Friedel oscillations (FO) is the quantum interference of electronic waves
scattering by the imperfection in crystalline host
materials\cite{Friedel1952}. Recently, STM has been demonstrated
experimentally to measure the topological Berry phase $\pi$ of monolayer
graphene by counting $2$ wavefront dislocations in Friedel oscillations, which
is induced by the intentional hydrogen adatom \cite{s41586-019-1613-5}. One
subsequent experiment shows that for bilayer graphene FOs can exhibit $4$,
$2$, or $0$ wavefront dislocations explained by the $2\pi$ Berry phase, the
specific sublattice positions of the single impurity and the position of STM
tip\cite{PhysRevLett.125.116804}. Electronic Berry phase as the intrinsic
nature of the wave functions is defined in momentum space, which is
responsible for many exotic electronic dynamics such as the index shift of
quantum Hall effect in monolayer graphene\cite{nature04233,nature04235} and
bilayer graphene\cite{nphys245}, Klein
tunneling\cite{nphys384,nphys1198} and the weak antilocalization
\cite{PhysRevLett.98.136801}. The probe of Berry phase usually requires the
magnetic field\cite{RevModPhys.82.1959,nature04233,nature04235,nphys1198}. These two
experiments not only do not need external magnetic field, but also realize the
measurement of Berry phase in real space. However, two experiments both focus
on the gapless cases for graphene and emphasize the relation between Berry
phase and wavefront dislocation number. This attracts us to explore the effect
of gap opening on the interference pattern of FOs.

Gap opening in graphene occurs in various different ways with the substrate
coupling as a typical example\cite{s41699-020-00162-4}. Most experiments and
devices are performed on the substrate-supported graphene, in which lattice
mismatch induced inversion symmetry breaking makes Berry phase unquantized
multiple of $\pi$\cite{YaoXiaoNiu2008}. As a result, gap opening should
challenge the established correspondence relation between Berry phase and
wavefront dislocation number in previous
experiments\cite{s41586-019-1613-5,PhysRevLett.125.116804}. In this study, we
study the FOs in gapped graphene. Comparing to the gapless graphene, gapped
graphene does not have pseudospin-momentum locking, this may prohibit the
occurrence of characteristic interference structure (namely, wavefront
dislocations) in FOs following the intuitive picture as suggested by the
seminal work \cite{s41586-019-1613-5}. But the wavefront dislocations in FOs
do emerge. Here, we explain the origin of wavefront dislocations by the
invariant pseudospin winding number in gaped and gapless graphene, which can
be regarded as an updated correspondence relation compatible with the previous
experiments\cite{s41586-019-1613-5,PhysRevLett.125.116804}. The wavefront
dislocations in FOs of gapped graphene can be verified in the
present experimental conditions, and this study helps deepen the understanding
of topological physics reflected in the FOs of the impurity-design system.


\begin{figure}[ptbh]
\includegraphics[width=1.0\columnwidth,clip]{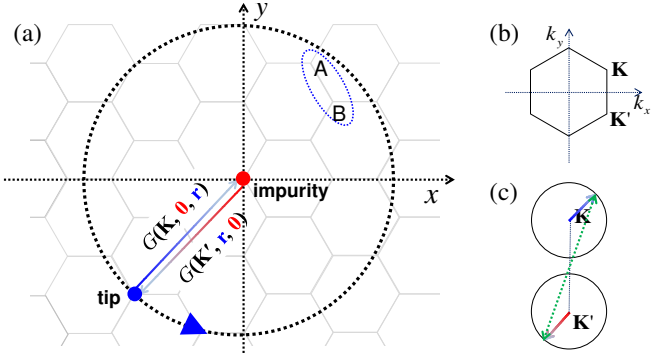}\caption{Schematic
measurement of electronic density oscillation by STM. (a) In real space atomic
structure of graphene, there is a single-atom vacancy (red dot) on one
sublattice site, by scanning the STM tip (blue dot) one can probe the
vacancy-induced density oscillations which are contributed by the intravalley
and intervalley scattering. Focusing on the intervalley scattering, we plot
corresponding physical process, i.e., STM tip emits the electron waves in one
valley $\mathbf{K}$, as shown by the propagator $G(\mathbf{K},\mathbf{0,r})$,
scattering by the vacancy back to STM tip through the other valley
$\mathbf{K}^{\prime}$, as shown by the propagator $G(\mathbf{K}^{\prime
},\mathbf{0,r})$. When the STM tip is shifted around the vacancy in real space
as shown by the blue arrow along the circle, the contributing momentum states
to the propagator change their momentum around $\mathbf{K}$ and $\mathbf{K}%
^{\prime}$ as shown in (c). (b) The Brillouin zone used to define $\mathbf{K}$
and $\mathbf{K}^{\prime}$. (c) Intervalley scattering (green double-arrowed
line) in momentum space. The contributing momentum states in each valley are
parallel to $\mathbf{r}$ and are denoted by using the same arrowed lines in
(a).}%
\label{scattering}%
\end{figure}

\begin{figure*}[ptbh]
\includegraphics[width=2.0\columnwidth,clip]{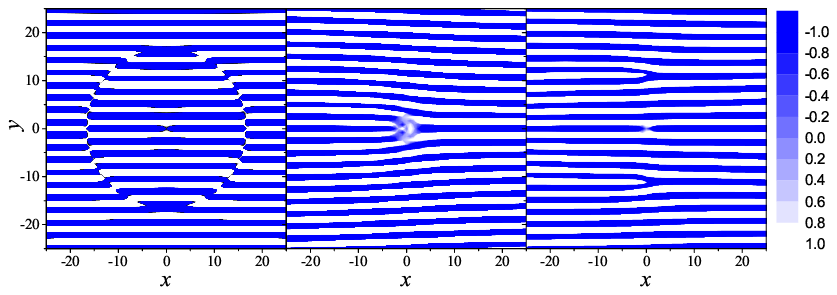}\caption{Friedel
oscillations pattern around a single-atom vacancy at the origin. Intervalley
scattering contribution to density oscillations $\delta\rho_{A}(\mathbf{r,}%
\varepsilon)$ and $\delta\rho_{B}(\mathbf{r,}\varepsilon)$ on sublattice $A$
(left panel) and on sublattice $B$ (middle panel), respectively. And the sum
$\delta\rho_{A}(\mathbf{r,}\varepsilon)+\delta\rho_{B}(\mathbf{r,}%
\varepsilon)$ as total electronic density modulation (right panel). Referring
to \onlinecite{s41586-019-1613-5}, the energy is integrated over energy up to
Fermi energy. Here, Fermi level $\varepsilon=0.11t_{0}\simeq0.4$ eV,
$\Delta=0.017t_{0}\simeq50$ mev, the color scale is normalized by a numerical
factor $C\times10^{-3}$.}%
\label{density}%
\end{figure*}

Fig. \ref{scattering} gives the schematic measurement of impurity-induced
electronic density oscillation of monolayer graphene by STM. The present
experimental technology allows one to intentionally introduce a single-atom
vacancy on arbitrary sublattice site of graphene\cite{437.full}, e.g., on the
sublattice $A$ as shown by the red dot in Fig. \ref{scattering}(a).
Corresponding to the introduction of the single vacancy, FOs occur, and lead
to the change of the space-resolved and energy-resolved local density of
states (LDOS) $\delta\rho(\mathbf{r,}\varepsilon)$%
\cite{BenaPRL2008,ZouPRB2016}:%

\begin{equation}
\delta\rho(\mathbf{r,}\varepsilon)=-\frac{1}{\pi}\operatorname{Im}%
[\text{Tr}\delta\mathbf{G}(\mathbf{r},\mathbf{r,}\varepsilon)], \label{LDOS}%
\end{equation}
where $\delta\mathbf{G}=\mathbf{G-G}^{0}$ represents\ the change of the total
Green's function (GF) or propagator $\mathbf{G}$ incorporating the effect of
the vacancy relevant to the bare propagator $\mathbf{G}^{0}$ of host system
(i.e., graphene in Fig. \ref{scattering}(a)), and has the form \
\begin{equation}
\delta\mathbf{G}(\mathbf{r}_{2},\mathbf{r}_{1}\mathbf{,}\varepsilon
)=\mathbf{G}^{0}(\mathbf{r}_{2},\mathbf{0,}\varepsilon)\mathbf{TG}%
^{0}(\mathbf{0},\mathbf{r}_{1},\varepsilon). \label{DeltGF}%
\end{equation}
Here, the $T$-matrix approach is used to describe the effect of vacancy whose
potential is simulated by $V_{0}\delta(\mathbf{r})$, and $T$-matrix
is\cite{PhysRevB.93.035413}
\begin{equation}
\mathbf{T(}\varepsilon\mathbf{)}=\mathbf{V}[1-\int d^{2}\mathbf{kG}%
^{0}(\mathbf{k,}\varepsilon)\mathbf{V}]^{-1}. \label{tmatrix}%
\end{equation}
In the $T$-matrix, $\mathbf{V}$ is usually a matrix and its form depends on
the specific position of the vacancy, e.g., in Fig. \ref{scattering}(a) (for the other vacancy configuration in supplementary materials), it is%

\begin{equation}
\mathbf{V}=\left[
\begin{array}
[c]{cc}%
V_{0} & 0\\
0 & 0
\end{array}
\right]  . \label{Vp}%
\end{equation}
For graphene, there are two Dirac valleys in the Brillouin
zone\cite{RevModPhys.81.109} (cf. Fig. \ref{scattering}(b)),  $\mathbf{K}=(\frac{2\pi}{3},\frac{2\pi}{3\sqrt{3}})$ and $\mathbf{K}^{\prime}=(\frac{2\pi}{3},-\frac{2\pi}{3\sqrt{3}})$, then $\delta\rho(\mathbf{r,}\varepsilon)$ are contributed by the intravalley
and intervalley scattering. In graphene, the intravalley scattering
contribution to $\delta\rho(\mathbf{r,}\varepsilon)$ has been well understood
through many
theoretical\cite{CheianovPRL2006,BenaPRL2008,HwangPRL2008,PhysRevB.78.014201,PhysRevB.79.125427,PhysRevB.80.094203,PhysRevB.82.193405,PhysRevLett.106.045504,PhysRevB.88.205416,PhysRevB.90.035440,PhysRevB.91.125408,PhysRevB.93.035413,PhysRevB.97.205410}
and
experimental\cite{219.full,PhysRevLett.101.206802,PhysRevB.86.045444,PhysRevX.4.011021,437.full}
efforts, while intervalley scattering contribution attracts
people's\ attention very recently due to its underlying topological
nature\cite{s41586-019-1613-5,PhysRevLett.125.116804,2008.03956}. $\delta
\rho(\mathbf{r,}\varepsilon)$ is measured easily by STM. FOs of $\delta
\rho(\mathbf{r,}\varepsilon)$ are dominated by the backscattering events along
the constant energy contour\cite{1764,jmmm.2019.165631}. Focusing on the
intervalley scattering, the corresponding physical process of Eq. \ref{DeltGF}
is shown in Fig. \ref{scattering}(a), i.e., STM tip emits the electron waves
from one valley $\mathbf{K}$, as shown by the propagator $G^{0}(\mathbf{K}%
,\mathbf{0,r,}\varepsilon)$, scattering by the vacancy back to STM tip through
the other valley $\mathbf{K}^{\prime}$, as shown by the propagator
$G^{0}(\mathbf{K}^{\prime},\mathbf{0,r,}\varepsilon)$. Of course, the
conjugate process also exists, i.e., the emission/scattering waves are from
$\mathbf{K}^{\prime}/\mathbf{K}$ valley. Here, $G^{0}(\mathbf{K/K}^{\prime
},\mathbf{0,r,}\varepsilon)$ is the matrix element of $2\times2$ matrix
$G^{0}$ and is associated with the valley momentum index. When the STM tip is
shifted around the vacancy in real space as shown in Fig. \ref{scattering}(a),
the contributing momentum states to the propagator change their momentum
around $\mathbf{K}$ and $\mathbf{K}^{\prime}$ as shown in Fig.
\ref{scattering}(c). As a result, the Berry phase defined in momentum space is
measured by STM in real space, and the key is the pseudospin-momentum
locking\cite{s41586-019-1613-5}. In contrast, we will show that the pseudospin
winding number instead of Berry phase is measured by STM in gaped graphene
without pseudospin-momentum locking.

The physics is essentially the same for gapped monolayer and bilayer graphene,
described below using gapped monolayer as an example and the relevant results for gapped bilayer graphene in supplementary materials.  The Hamiltonian of gaped monolayer graphene is $H_{0}=v_{F}(\eta\sigma_{x}k_{x}%
-\sigma_{y}k_{y})+\Delta\sigma_{z}$. $H_{0}$ is expressed in the sublattice
basis of $\{A,B\}$, then $\sigma_{x,y,z}$ is the Pauli matrix acting on the
pseudospin space. $\pm\Delta$ is the staggered potential on the sublattice $A$
and $B$, which originates from the inversion symmetry breaking, e.g., by the
proximity substrate\cite{nmat2003,s41699-020-00162-4}. And $\eta=\pm1$ is
valley index for two inequivalent valleys in graphene, $v_{F}=3/2a_{0}t_{0}$
with $a_{0}$ being the carbon-carbon bond length and $t_{0}$ being the
nearest-neighbor hopping energy, and $a_{0}$/$t_{0}$ is used as\ the
length/enegy unit in our convention\cite{RevModPhys.81.109}. The energy
spectrum and the spinor wavefunction are, respectively,

\begin{equation}
E_{\xi}(\mathbf{k})=\xi v_{F}\sqrt{\delta^{2}+k^{2}},
\end{equation}
and%
\begin{equation}
\psi_{\xi,\mathbf{k}}=\frac{1}{\sqrt{1+k^{2}/(\epsilon_{\xi}+\delta)^{2}}%
}\left[
\begin{array}
[c]{c}%
1\\
\frac{\eta k_{x}-ik_{y}}{\epsilon_{\xi}+\delta}%
\end{array}
\right]  ,
\end{equation}
where the reduced quantities are defined as $\delta=\Delta/\nu_{F}$,
$\epsilon_{\xi}=E_{\xi}/v_{F}$ with $\xi=\pm1$ for the conduction and the
valence band. The GF in momentum space is defined as $G^{0}\left(
\mathbf{k,}\varepsilon\right)  \equiv(z-H_{0})^{-1}$, and it is
\begin{equation}
\mathbf{G}^{0}\left(  \mathbf{k,}\varepsilon\right)  =\frac{1}{z^{2}%
-\Delta^{2}-k^{2}\nu_{F}^{2}}\left[
\begin{array}
[c]{cc}%
z+\Delta & \eta v_{F}ke^{i\eta\theta_{k}}\\
\eta v_{F}ke^{-i\eta\theta_{k}} & z-\Delta
\end{array}
\right]  .
\end{equation}
Here, $z=\varepsilon+i0^{+}$ with $0^{+}$ for the retarded properties of GF and the Fermi level $\varepsilon$ is assumed in the conduction band for brevity.
Performing the Fourier transformation to the momentum space GF, we express the
real space GF\cite{ZhuPRL2011} as $\mathbf{G}^{(0)}(\mathbf{K},\mathbf{r,}%
\varepsilon)=-e^{i\mathbf{K}\cdot\mathbf{r}}/(2v_{F})^{2}\mathbf{G}%
^{(0)}(\mathbf{r,}\varepsilon)$ with%

\begin{equation}
\mathbf{G}^{(0)}(\mathbf{r,}\varepsilon)=\left[
\begin{array}
[c]{cc}%
i\varepsilon_{+}H_{0}\left(  u\right)  & \eta
\sqrt{\varepsilon_{+}\varepsilon_{-}}H_{1}\left(
u\right)  e^{i\eta\theta_{r}}\\
\eta\sqrt{\varepsilon_{+}\varepsilon_{-}}H_{1}\left(
u\right)  e^{-i\eta\theta_{r}} & i\varepsilon_{-}H_{0}\left(  u\right)
\end{array}
\right]  ,
\end{equation}
where $H_{j}$ is $j$-th order Handel function of the first kind, $\varepsilon_{\pm}=\varepsilon\pm\Delta$, $u=r\sqrt{\varepsilon
_{+}\varepsilon_{-}}/v_{F}$, and $r$ ($\theta_{r})$ is the module (azimuthal
angle) of $\mathbf{r}$\textbf{.} Concentrating on the intervalley
contribution, the change of LDOS is%

\begin{equation}
\delta\rho(\Delta\mathbf{K,r},\varepsilon)=\delta\rho_{A}(\mathbf{r}%
,\varepsilon)-\delta\rho_{B}(\mathbf{r},\varepsilon), \label{interLDOS}%
\end{equation}
where the sublattice-resolved LDOS are%

\begin{subequations}
\begin{align}
\delta\rho_{A}(\mathbf{r},\varepsilon)  &  =C\operatorname{Im}\left[
t(\varepsilon)H_{0}^{2}(u)\varepsilon_{+}^{2}\right]  \cos(\Delta
\mathbf{K}\cdot\mathbf{r}),\\
\delta\rho_{B}(\mathbf{r},\varepsilon)  &  =C\operatorname{Im}\left[
t(\varepsilon)H_{1}^{2}(u)\varepsilon_{+}\varepsilon_{-}\right]  \cos
(\Delta\mathbf{K}\cdot\mathbf{r}-\Delta\eta\theta_{r}).
\end{align}
Here, $\Delta\mathbf{K=K-K}^{\prime}$, $\Delta\eta=\eta-\eta^{\prime}%
$\textbf{,}$\ C=1/(8v_{F}^{4})$ and $t(\varepsilon)=V_{0}/[1-V_{0}G_{AA}%
^{(0)}(0,\varepsilon)]$ is the matrix element of $T$-matrix induced by the
vacancy on the sublattice $A$.

\begin{figure}[ptbh]
\includegraphics[width=1.0\columnwidth,clip]{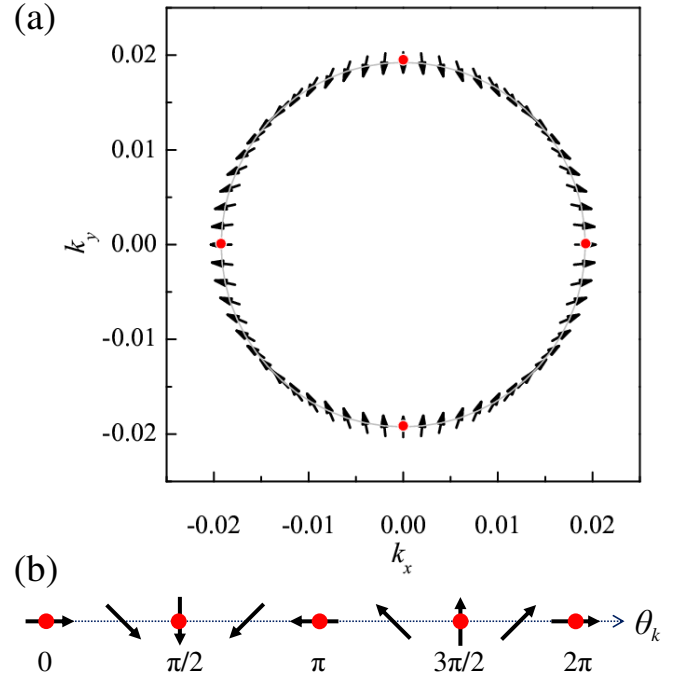}\caption{Pseudospin
texture pattern. (a) Pseudospins on the constant energy contour. (b)
Pseudospin as the function of $\theta_{k}$. Here, $\Delta=50$ meV,
$\varepsilon=100$ meV and $\eta=1$.}%
\label{texture}
\end{figure}

Nontrivially, Eq. \ref{DeltGF} for the LDOS contributed by the intervalley
scattering has the identical form as that in gapless graphene
\cite{s41586-019-1613-5} and it reproduces the result of gapless case when
$\Delta\rightarrow0$. For the vacancy on the sublattice $A$, $\delta\rho
_{A}(\mathbf{r},\varepsilon)$ is trivial. $\left\vert \Delta\eta\right\vert
=2$ for the intervalley scattering, then the phase of $\delta\rho
_{B}(\mathbf{r},\varepsilon)$ is singular at $\mathbf{r}=0$. By shifting STM
tip around the vacancy, i.e., $\theta_{r}$ is rotated by $2\pi$, there should
be $2$ additional wavefronts in the Friedel oscillation pattern of $\delta
\rho_{B}(\mathbf{r},\varepsilon)$\cite{s41586-019-1613-5}. In Fig.
\ref{density}, focusing on the intervalley scattering contribution, we show
electronic density oscillations around a single-atom vacancy $\delta\rho
_{A}(\mathbf{r},\varepsilon)$ and $\delta\rho_{B}(\mathbf{r},\varepsilon)$ on
sublattice $A$ (left panel) and on sublattice $B$ (middle panel),
respectively. And the sum $\delta\rho_{A}(\mathbf{r},\varepsilon)+\delta
\rho_{B}(\mathbf{r},\varepsilon)$ as total electronic density modulation
(right panel). As expected, $\delta\rho_{A}(\mathbf{r},\varepsilon)$ exhibit
the normal oscillating wavefronts perpendicular to $\Delta\mathbf{K}$ with a
wavelength $\lambda_{\Delta\mathbf{K}}=2\pi/\left\vert \Delta\mathbf{K}%
\right\vert =2.60a_{0}\approx3.69$ $\mathring{a}$, and does not display any
topological feature. $\delta\rho_{B}(\mathbf{r},\varepsilon)$ gives two
wavefront dislocations at $\mathbf{r}=0$, which accommodates for the phase
accumulated along the contour enclosing the singular point of the phase
$\theta_{r}$\cite{rspa.1974.0012}. In the total electronic density modulation
(cf. right panel of Fig. \ref{density}), $\delta\rho_{A}(\mathbf{r}%
,\varepsilon)$ only shifts the position of dislocations from $\mathbf{r}=0$
along the direction parallel to $\Delta\mathbf{K}$\textbf{, }but does not
change the shape and the number of dislocations\cite{s41586-019-1613-5}.

The number $2$ of additional wavefronts is regarded as the signature of the
Berry phase $\pi$ of graphene. However, the correspondence between Berry phase
and the wavefront dislocation number fails since the Berry phase is
unquantized multiple of $\pi$ in gapped graphene. Back to Fig.
\ref{scattering}(c), in which we do not follow the
references\cite{s41586-019-1613-5,PhysRevLett.125.116804} to plot the
momentum-resolved pseudospin direction, but it still shows synchronous motion
of the dominant scattering state contributing to FOs in momentum space and STM
tip in real space, e.g., the scattering state rotates
clockwise/counterclockwise\ on the constant energy contour when STM tip is
shifted clockwise/counterclockwise around the vacancy. The scattering state
can be used to define the pseudospin vector $\mathbf{s=(}s_{x},s_{y}%
\mathbf{)}$ where%

\end{subequations}
\begin{subequations}
\label{spin}%
\begin{align}
s_{x}  &  =\left\langle \psi_{\xi,\mathbf{k}}\left\vert \sigma_{x}\right\vert
\psi_{\xi,\mathbf{k}}\right\rangle =\frac{2\eta k_{x}(\epsilon_{\xi}+\delta
)}{(\epsilon_{\xi}+\delta)^{2}+k^{2}},\\
s_{y}  &  =\left\langle \psi_{\xi,\mathbf{k}}\left\vert \sigma_{y}\right\vert
\psi_{\xi,\mathbf{k}}\right\rangle =-\frac{2k_{y}(\epsilon_{\xi}+\delta
)}{(\epsilon_{\xi}+\delta)^{2}+k^{2}}.
\end{align}
Due to $s_{x}\sim k_{x}$ and $s_{y}\sim k_{y}$, Eq. \ref{spin} gives $4$
fixing points with the momentum azimuthal angle $\theta_{k}=0$, $\pi/2$, $\pi
$, and $2\pi$, at which the pseudospin directions are fixed. This feature has
no dependence on $\delta$, then is robust to the gap opening. To show more
visually, we arbitrarily choose a set of parameters to plot the pseudospin
texture in Fig. \ref{texture}. With the scattering state moving on the
constant energy contour, the pseudospin direction twists continuously in Fig.
\ref{texture}(a). Clearly, there are $4$ fixing points shown by the red dots
(cf. Fig. \ref{texture}(a)), which implies the invariant pseudospin winding
number. Referring to Fig. \ref{texture}(b), when $\theta_{k}$ evolves from $0$
to $2\pi$, the pseudospin rotates by $2\pi$ corresponding to the winding
number $1$. Alternatively, the invariant pseudospin winding number can also be
shown mathematically\cite{2010-00259-2}. To rewrite the gapped Hamiltonian of monolayer graphene
into the form:%

\end{subequations}
\begin{equation}
H_{0}(\mathbf{k})=\left\vert E_{\xi}(\mathbf{k})\right\vert \left[
\begin{array}
[c]{cc}%
\cos\alpha & \sin\alpha e^{-i\phi}\\
\sin\alpha e^{i\phi} & -\cos\alpha
\end{array}
\right]  ,
\end{equation}
where the azimuthal $\alpha(\mathbf{k})$ and polar $\phi(\mathbf{k})$ angles
on the Bloch sphere are defined as $\cos\alpha=\Delta/\left\vert E_{\xi
}(\mathbf{k})\right\vert $, $\sin\alpha=f(\mathbf{k})/\left\vert E_{\xi
}(\mathbf{k})\right\vert $ with $f(\mathbf{k})=v_{F}(\eta k_{x}+ik_{y})$ and
$\phi\equiv-$Arg$f$. As a result, we obtain the Berry phase along a closed
Fermi surface:%

\begin{equation}
\gamma(C)=%
{\displaystyle\oint\limits_{C}}
d\mathbf{k\cdot A_{\xi}}=\pi W_{C}[1-\frac{\Delta}{\left\vert E_{\xi
}(\mathbf{k})\right\vert }].\label{Berry}
\end{equation}
Here, we have used the Berry connection $\mathbf{A_{\xi}}\mathbf{=}%
i\left\langle u_{\mathbf{k},\xi}|\nabla_{\mathbf{k}}u_{\mathbf{k},\xi
}\right\rangle =-\xi\sin^{2}\frac{\alpha}{2}\nabla_{\mathbf{k}}\phi$. Most
importantly, the winding number is introduced
\begin{equation}
W_{C}\equiv-\xi%
{\displaystyle\oint\limits_{C}}
\frac{d\phi}{2\pi}=\eta\xi. \label{Winding}
\end{equation}
While the Berry phase of Eq. \ref{Berry} has the $\Delta$ dependence and becomes unquantized
multiple of $\pi$, $W_{C}$ of Eq. \ref{Winding} with the absolute value $1$ is topologically
invariant in gapless and gapped graphene. As a result, the $2$ wavefront
dislocations also exist in gapped graphene, which should be ascribed to the
invariant winding number.


Here, we firstly discuss the experimental implication of our theoretical discovery.
When $\Delta\rightarrow0$ for gapless graphene, one can say that the Berry
phase and winding number are equivalent, then Berry phase is
measured by the wavefront dislocations\cite{s41586-019-1613-5,PhysRevLett.125.116804}. But a supplementary experiment is
necessary to confirm the gapless nature of graphene sample. In previous two
experiments\cite{s41586-019-1613-5,PhysRevLett.125.116804}, there is no such supplementary experiment and they do
not perform the quantitative comparison between experiments and simulations
for the FOs, which both prohibit the discovery of the possible gap opening. The quantitative
calculations of FOs, especially properly incorporating the strength and range of impurity potential, is rather important and is worth of further
simulations, e.g., by using density functional calculations\cite{PhysRevB.102.195416}. Also, this experimental discussions are also applicable to one recent theoretical proposal, which uses the presence or absence of wavefront dislocations in FOs to distinguish the incompatible low-energy models for twisted bilayer graphene\cite{PhysRevLett.125.176404}. In light of present experimental technology, we expect the immediate verification of our theoretical prediction in gapped graphene. In addition, our results may stimulate the theoretical and experimental research interest on the FOs in intrinsic gapped two-dimensional materials, e.g., transition metal dichalcogenides\cite{nnano.2012.193,natrevmats.2016.55}.

The robustness of the wavefront dislocations of FOs to the gap opening
dictates the potential applications in graphene-based pseudospintronics\cite{PesinNatMater2012}.
Each imperfection can be regarded as one vortex, and the sublattice dependence of wavefront dislocation help define the vortex and
antivortex since the orientations of the tripod shape of the imperfection on two sublattices are different\cite{s41586-019-1613-5}. Then, we can use the vortex/antivortex as 0/1 to construct the memory device.
And the the writing of vortex-based memory device can be performed by using STM to shift the imperfection from one sublattice to the other one\cite{437.full}, while its reading is realized by scanning the imperfection-induced FOs to characterize the orientations of the tripod shape of the imperfection. One recent experiment\cite{2008.03956} studies the effect of spacing on the wavefront dislocation of two vortices, and shows the irrelevant vortices on the several nanometer scale which favors the high-density of vortex-based memory device. This implies the great potential of impurity-design graphene in pseudospintronics.


In summary, we have studied the Friedel oscillation in gapped graphene.
Although the pseudospin-momentum locking is broken, the wavefront dislocations still emerge in intervalley scattering contributed Friedel
oscillations. We establish the correspondence between the invariant winding number instead of the Berry phase and the number of wavefront dislocations in gapped graphene, which can be verified by the present experimental technology. This study is helpful to the understanding of
correspondence between topological quantity and wavefront dislocations in
Friedel oscillations, and broadens the range of materials to observe the wavefront dislocations of FOs, e,g, in transition metal dichalcogenides\cite{nnano.2012.193,natrevmats.2016.55}.

\section*{Acknowledgements}

The authors thank Dr. Li-Kun Shi for helpful
discussions. This work was supported by the National Key R$\&$D Program of China (Grant No.
2017YFA0303400), the National Natural Science Foundation of China (NSFC)
(Grants No. 11774021 and No. 11504018), and the NSAF grant in NSFC (Grant No.
U1930402). We acknowledge the computational support from the Beijing
Computational Science Research Center (CSRC).


\end{document}